\documentclass[
 reprint,
 amsmath,amssymb,
 aps,
 pre   
]{revtex4-2}

\usepackage{silence}
\usepackage{savesym}
\savesymbol{doi}
\usepackage[utf8]{inputenc} 
\usepackage[T1]{fontenc}    
\usepackage{varioref}
\usepackage{url}            
\usepackage{booktabs}       
\usepackage{amsfonts}       
\usepackage{nicefrac}       
\usepackage{microtype}      
\usepackage{natbib}
\usepackage{doi}
\usepackage{amsmath}
\usepackage{xspace}
\usepackage{graphicx}
\usepackage{dcolumn}
\usepackage{bm}
\usepackage{hyperref}
\usepackage{cleveref}

\begin{document}

\preprint{APS/123-QED}

\title{\texorpdfstring{Quantifying the influence of Vocational Education and Training \\ with text embedding and similarity-based networks}{Quantifying the influence of Vocational Education and Training with text embedding and similarity-based networks}}%


\author{Hyeongjae Lee}
\affiliation{Graduate School of Data Science, Chonnam National University, Gwangju, Republic of Korea}
\author{Inho Hong}
\email{ihong@jnu.ac.kr}
\affiliation{Graduate School of Data Science, Chonnam National University, Gwangju, Republic of Korea}

\date{\today}

\begin{abstract}
Assessing the potential influence of Vocational Education and Training (VET) courses on creating job opportunities and nurturing work skills has been considered challenging due to the ambiguity in defining their complex relationships and connections with the local economy. Here, we quantify the potential influence of VET courses and explain it with future economy and specialization by constructing a network of more than 17,000 courses, jobs, and skills in Singapore's SkillsFuture data based on their text similarities captured by a text embedding technique, Sentence Transformer. We find that VET courses associated with Singapore's 4th Industrial Revolution economy demonstrate higher influence than those related to other future economies. The course influence varies greatly across different sectors, attributed to the level of specificity of the skills covered. Lastly, we show a notable concentration of VET supply in certain occupation sectors requiring general skills, underscoring a disproportionate distribution of education supply for the labor market. 
Secondary publications and information 
\end{abstract}

\maketitle



Vocational Education and Training (VET) enhances human capital development by directly connecting with technological standards, the organization of the production process, and culturally specific work divisions, all of which necessitate work-related upper-secondary and post-secondary qualifications
~\citep{wallenborn2010vocational}. From the individual perspective, VET not only facilitates smooth entry into the labor market early in an employee’s career but also supports long-term career stability by enhancing adaptability to rapid technological and structural changes in the economy through reskilling and upskilling~\citep{vinayan2020upskilling, li2022reskilling, pradhan2023reskilling}. 

Despite its importance in the labor market, quantifying the influence of VET and assessing its effectiveness remains challenging due to its complex relationships with local environments. Unlike a standardized or global model, VET often functions as a subsystem shaped by stimuli at the country or regional level~\citep{han2023supply, moso2022quantitative, greinert2004european}. An intriguing indicator proposed to measure VET effectiveness is vocational or occupational specificity, focusing on evaluating VET in the context of School-to-Work linkage~\citep{souto2012coherence, eggenberger2018occupational}. While this index provides valuable insights, a practical challenge arises in profiling and tracking the education-to-employment pathways of individuals who have completed specific VET programs~\citep{masdonati2010vocational,brunetti2019school}. This challenge is particularly pronounced given the contemporary labor market trend where individuals' school-to-work transitions and career pathways throughout their lifespan are increasingly diversified~\citep{forster2018vocational, schoon2016diverse}.

As numerous countries are making substantial investments in VET to ensure that their labor forces are well-equipped with the necessary skills and readiness for emerging industries~\citep{ranasinghe2019school, brown2023skill}, another crucial aspect of VET programs worth investigating is an imbalance between supply and demand in the VET market~\citep{lopes2023supply}. A disparity between the supply and demand of VET provisions turns out to be a key factor in hindering educational equality, which is closely related to the development of a regional economy~\citep{choi2021impact}. In that sense, detecting any imbalance in the VET market of a region holds implications for informed policy-making in addressing educational challenges, leading to the sustainable development of the region~\citep{mcgrath2016skills,viertel2010vocational}.

Analyzing occupations, human capital, and VET courses as an interconnected network enables the examination of their relationships and the complementarity among various VET programs~\citep{budel2020complementarity}. Network analysis has been implemented for exploring the interconnections of human capital entities over different labor markets. Utilizing the co-occurrence of skills among job seekers and providers from O*NET data, researchers explained the economic inequality and the hollowing of the middle class with the polarization between cognitive and physical skills in high-wage workers and low-wage workers within the constructed networks~\citep{anderson2017skill}. Similarly, a collaboration network of workers with different educational backgrounds revealed that co-workers who are more synergistic are more likely to substitute for one another~\citep{neffke2019value}. As such, while the labor market and human capital domain have been actively explored, the VET domain largely remains unexplored. This gap arises from the practical issue of finding a suitable dataset that captures the direct connections between VET courses, occupations, and skills as individual entities. 

Recent advancements in Natural Language Processing (NLP) can provide an alternative approach to overcome this difficulty~\citep{matsui2022word, 10098736}. To be specific, contextual embedding using unstructured textual data from various sources has enabled researchers to access the semantic dimension of socio-economic entities and explore their relationships within high-dimensional vector space~\citep{mikolov2013efficient, liu2020survey}. For instance, trained on 850 billion words in English-language books from the Google n-grams dataset, historical shifts in different social-cultural groups are traced by comparing the top words associated with each group from the embedded space~\citep{charlesworth2022historical}. Likewise, utilizing a Google news dataset and a pre-trained w2vNEWs model, gender bias has been revealed through vector analogy between English vocabularies~\citep{bolukbasi2016man}. Generating a culture dictionary and scoring words based on word representation trained from earning call transcripts, a study has attempted to measure corporate culture and demonstrate its association with business outcomes~\citep{li2021measuring}. 

When it comes to exploring the semantic relations between entities in human capital and labor market research, the heterogeneous entities of industries, occupations, skills, and firms can be mapped onto a unified labor space using the recent BERT model~\citep{kim2023labor, devlin2018bert}. Additionally, by focusing on university course syllabi and measuring their distance from frontier knowledge within the embedded space, the innovation gap across higher educational institutions has been explained with the academic and economic outcomes of graduates~\citep{biasi2022education}. The recently publicized dataset relating higher education curricula and skills based on text embedding suggests a new possibility to quantify the relationship between VET courses and skills~\citep{javadian2024course}. These cases highlight the potential of NLP in extracting meaningful insights from unstructured textual data across diverse fields, showcasing its versatility and applicability in understanding complex dynamics in the education-labor market. 

To embed the entire text of each unit into a single vector, the study utilizes the Sen\-tence Transformer model in constructing a network based on the se\-man\-tic tex\-tu\-al sim\-i\-lar\-i\-ty~\citep{reimers2019sentence, chandrasekaran2021evolution, farouk2020measuring}, drawing on tex\-tu\-al data extracted from descriptions of VET course, skill, and oc\-cu\-pa\-tions provided by Sin\-ga\-pore's SkillsFuture initiative. 

Although not identically the same, there is a similar technique called Sem\-an\-tic Net\-work Analysis (SNA)~\citep{drieger2013semantic}. SNA is a technique to discover se\-man\-tic structures in texts by constructing a co-oc\-cur\-rence matrix and a net\-work based on the words in texts. Se\-man\-tic net\-work analysis has been applied to understand how user opinions and experiences are communicated within online communities or how different concepts are represented or related to each other~\citep{kang2017semantic, ruiz2015exploring}. While similar, this study differs in that it creates a co-oc\-cur\-rence matrix and net\-work based on the sim\-i\-lar\-i\-ty of VET/skill/oc\-cu\-pa\-tion units.

From this background, the study aims to answer two established labor market research questions: (1) measuring the influence of one entity over another and (2) elucidating the structure of entities within the constructed network~\citep{kim2023labor}. To accomplish this, our analysis of Singapore's data seeks to quantitatively assess the potential implications of each VET entity, introducing a novel index. Then, we discuss the fair distribution of VET course supply over different jobs by employing an occupational supply index.

\section*{Results}

\subsection*{Course network analysis}
\begin{figure*}[htbp]
    \centering
\includegraphics[width=\textwidth]{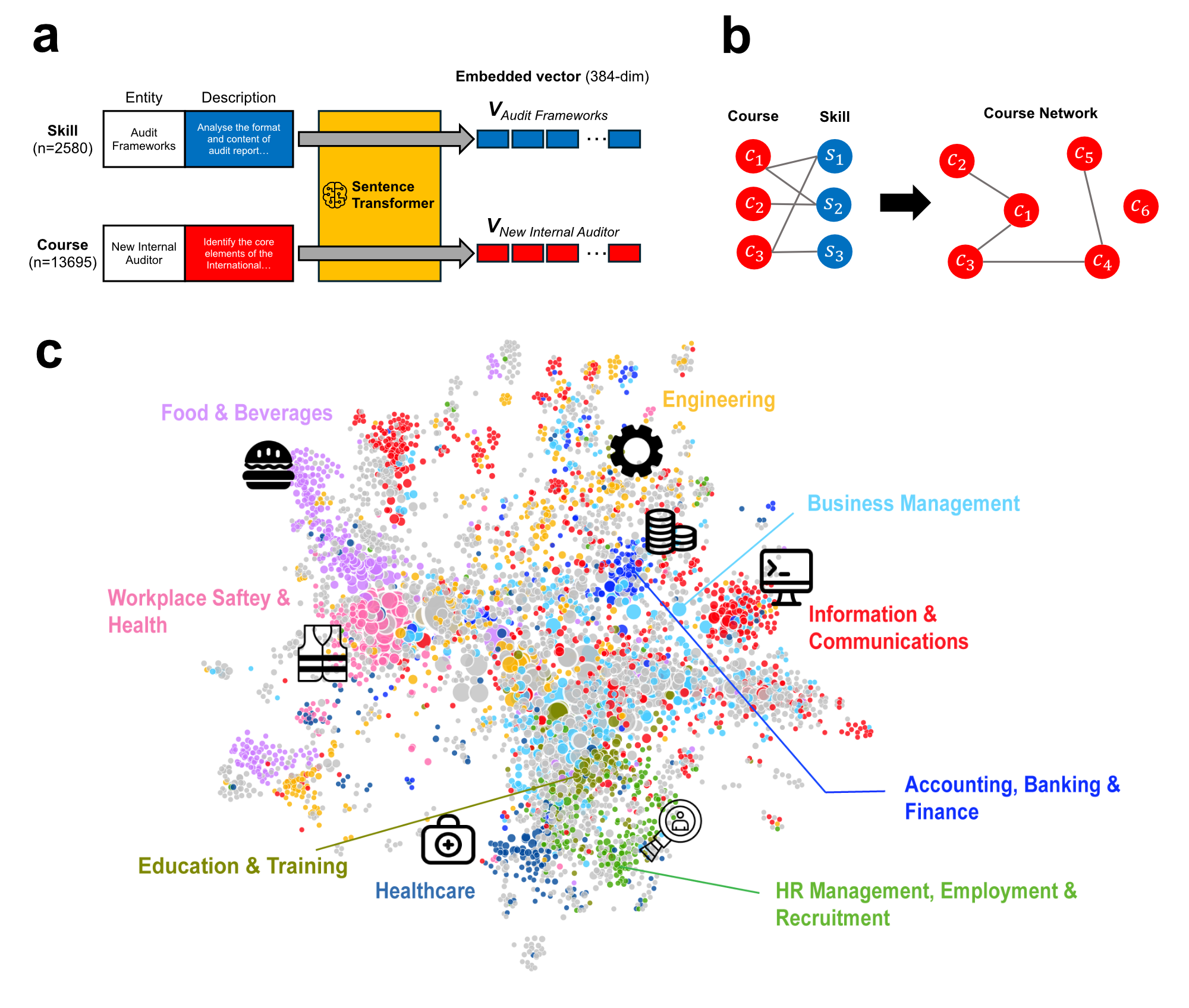}
    \caption{Schematic of text embedding. {\bf a}. The diagram shows the embedding process, where textual descriptions of skills and VET courses are transformed into vector representations using a pre-trained Sentence Transformer model. {\bf b}. Schematic of constructing a course network from the course-skill bipartite network. The course network is the co-skill network of VET courses. {\bf c}. Illustration of the course network. Each node represents a VET course that has more than one neighbor from the course-skill bipartite network. The node color signifies the VET course category, while the node size demonstrates the degree centrality. Links are omitted in the visualization for simplicity. The colored nodes represent the top 8 categories with the highest number of VET courses, highlighting the most dominant sectors in SkillsFuture Singapore.}
    \label{fig1}
\end{figure*}

To illustrate the linkage between VET courses through their associated work skills, we first constructed the VET course network by projecting the bipartite network of courses and skills onto a monopartite course network. The projected VET course network successfully preserves the textual attributes of the analyzed entities, maintaining relationships among skill and course entities. The clustering of courses by their categories highlights how the network retains the characteristics of entities (see Fig~\ref{fig1}). The finding can be further backed up by the average path length between the top 10 VET sectors, which reflects the proximity between related sectors (see Fig~\ref{S1_Fig}). For example, sectors associated with human management, such as education and HR, are closely interconnected. Likewise, the business management sector exhibits relative proximity to other relevant sectors such as sales, marketing, and information technology. 

\begin{figure*}[htbp]
\centering
\includegraphics[width=1\textwidth]{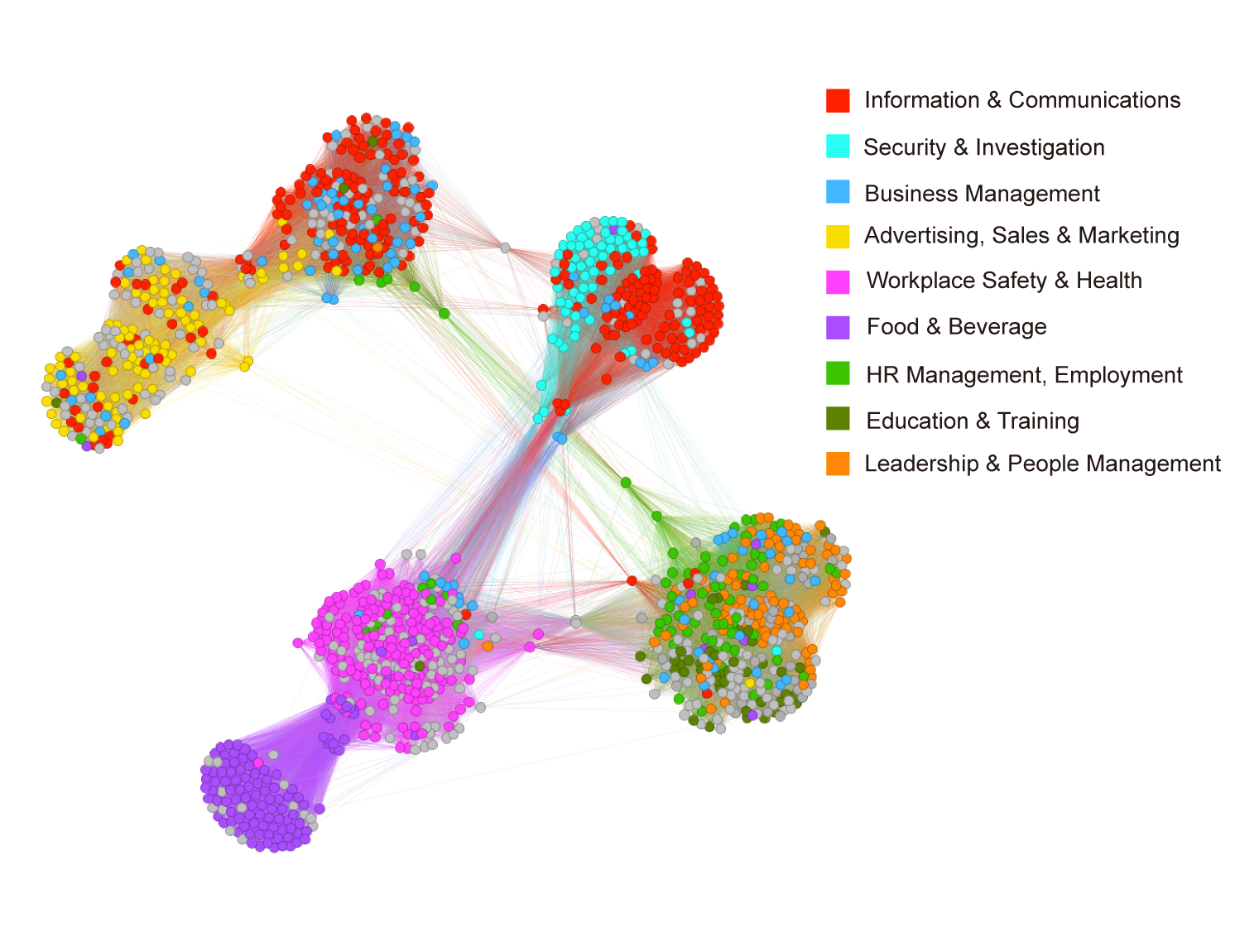}
\caption{{The $k-$core subgraph with $k=100$.}
The color of the nodes indicates different VET course sectors.}
\label{fig2}
\end{figure*}

The $k$-core decomposition, removing nodes with fewer connections than the specified $k$~\citep{batagelj2003m}, supports the presence of distinct clusters represented by the course categories.
In Fig.~\ref{fig2}, six separate clusters are found for $k=100$, representing (1) sales, (2) information, (3) mix of information and security, (4) safety \& health, (5) Food, and (6) HR and management.

\subsection*{Measuring course influence in Singapore VET market}

We propose a metric called ``course influence'' to quantify the potential implication of individual VET courses. The metric aims to assess the potential influence of a VET course on work skills given the number of similar courses that could provide duplicated education content. On the one hand, the coverage \( C_c \) of a course $c$ counts the number of connected skills in the course-skill network, which represents the abilities for employees to acquire for school-job transition or career development, thereby reflecting labor market demand. On the other hand, the degree centrality (\( k_c \)) of a course, counting the number of other courses with similar descriptions, denotes the complementarity between VET courses, indicating the supply within the educational market. Then, the course influence $CI_{c}$ is calculated by dividing each course's coverage by its degree centrality as

\begin{eqnarray}
\label{eq:course_influence}
CI_c &=& \frac{C_c}{k_c}.
\end{eqnarray}

The index exhibits an exponential distribution for our dataset (see Fig~\ref{S2_Fig}
)

\subsection*{Course influence and future economy}

The suggested course influence index of VET courses needs to be explained with respect to their diversity in targeted skills since determining the types and number of skills is an essential step for designing curricula and course content. We incorporated VET skill diversity and VET transferability in our regression model to examine their relationship with the course influence of VET courses. We constructed Ordinary Least Squares (OLS) regression models with course influence as the dependent variable to delve deeper into this question. Model (1) integrates variables for the diversity and transferability associated with four distinct future economies as independent variables. Model (2) utilizes only those variables identified as significant from model (1). Model (3) substitutes the diversity factor with salary to examine the relationship between salary and course influence. The average salary for each course was calculated as the average monthly income from related occupations within the course-skill network. Lastly, Model (4) includes the same variables as the previous models but excludes the information of VET sectors to compare the results with and without this control.

The regression models reveal two key factors associated with course influence: transferability for the Fourth Industrial Revolution and low skill diversity. From model (1), only the transferability value related to the Fourth Industrial Revolution economy is positively associated with course influence (0.1538***) among the four transferability indices. This indicates that VET courses related to skills creating job opportunities for the Fourth Industrial Revolution economy tend to have greater potential implications by covering more skills effectively. In contrast, the transferability for both digital and care economies is negatively associated with course influence (-0.1015*** and –0.0611**, respectively), suggesting that courses related to these economies tend to cover fewer work skills. 
Additionally, the negative coefficient for skill diversity (-0.5012***) indicates that courses spanning different skill sectors, i.e., covering more general skills, are likely to cover fewer skills.

Model (2), which includes only the highly significant variables from the previous model, confirms the robustness of Model (1). Model (3) investigates the relationship between average salary and course influence, revealing that the average salary does not have a significant relationship with course influence. Finally, Model (4) repeats the analysis without controlling for VET sectors, showing the robustness of the results for this control.

\begin{table*}[!htbp]\centering 
\caption{Regression analysis on VET course influence. VET course labels are under control within the model.}
  \label{Tab1} 
\renewcommand{\arraystretch}{1.2} 
\begin{tabular}{@{\extracolsep{5pt}}l p{2cm} p{2cm} p{2cm} p{2cm} p{2cm} p{2cm} p{2cm} p{2cm}} 
\\[-1.8ex]\hline 
\hline \\[-1.8ex] 
& \multicolumn{8}{c}{\textit{Dependent variable: Course Influence}} \\ 
\cline{2-9} 
\\[-1.8ex] & \multicolumn{2}{c}{(1)} & \multicolumn{2}{c}{(2)} & \multicolumn{2}{c}{(3)} & \multicolumn{2}{c}{(4)} \\ 
\hline \\[-1.8ex] 

Intercept & \multicolumn{2}{c}{0.6598$^{***}$} & \multicolumn{2}{c}{0.6561$^{***}$} & \multicolumn{2}{c}{0.2051$^{***}$} & \multicolumn{2}{c}{0.6100$^{***}$} \\ 
 & \multicolumn{2}{c}{(0.047)} & \multicolumn{2}{c}{(0.047)} & \multicolumn{2}{c}{(0.019)} & \multicolumn{2}{c}{(0.045)} \\ 

Transferability (4th IR) & \multicolumn{2}{c}{0.1538$^{***}$} & \multicolumn{2}{c}{0.1480$^{***}$} & \multicolumn{2}{c}{0.1024$^{***}$} & \multicolumn{2}{c}{0.1803$^{***}$} \\ 
 & \multicolumn{2}{c}{(0.033)} & \multicolumn{2}{c}{(0.033)} & \multicolumn{2}{c}{(0.034)} & \multicolumn{2}{c}{(0.032)} \\ 

Transferability (Digital) & \multicolumn{2}{c}{$-$0.1015$^{***}$} & \multicolumn{2}{c}{$-$0.1106$^{***}$} & \multicolumn{2}{c}{$-$0.1024$^{***}$} & \multicolumn{2}{c}{$-$0.0857$^{***}$} \\ 
 & \multicolumn{2}{c}{(0.028)} & \multicolumn{2}{c}{(0.028)} & \multicolumn{2}{c}{(0.029)} & \multicolumn{2}{c}{(0.026)} \\ 

Transferability (Care) & \multicolumn{2}{c}{$-$0.0611$^{**}$} & \multicolumn{2}{c}{} & \multicolumn{2}{c}{$-$0.0595$^{**}$} & \multicolumn{2}{c}{$-$0.1196	$^{**}$} \\ 
 & \multicolumn{2}{c}{(0.029)} & \multicolumn{2}{c}{} & \multicolumn{2}{c}{(0.030)} & \multicolumn{2}{c}{(0.027)} \\ 

Transferability (Green) & \multicolumn{2}{c}{$-$0.0351} & \multicolumn{2}{c}{} & \multicolumn{2}{c}{$-$0.0440} & \multicolumn{2}{c}{$-$0.0275} \\ 
 & \multicolumn{2}{c}{(0.045)} & \multicolumn{2}{c}{} & \multicolumn{2}{c}{(0.049)} & \multicolumn{2}{c}{(0.045)} \\ 

Skill Diversity & \multicolumn{2}{c}{$-$0.5012$^{***}$} & \multicolumn{2}{c}{$-$0.4992$^{***}$} & \multicolumn{2}{c}{} & \multicolumn{2}{c}{$-$0.5067	$^{***}$} \\ 
 & \multicolumn{2}{c}{(0.048)} & \multicolumn{2}{c}{(0.048)} & \multicolumn{2}{c}{} & \multicolumn{2}{c}{(0.047)} \\ 

Average Salary & \multicolumn{2}{c}{} & \multicolumn{2}{c}{} & \multicolumn{2}{c}{$-$0.0090} & \multicolumn{2}{c}{} \\ 
 & \multicolumn{2}{c}{} & \multicolumn{2}{c}{} & \multicolumn{2}{c}{(0.029)} & \multicolumn{2}{c}{} \\ 

VET sector FE (fixed effect) & \multicolumn{2}{c}{Yes} & \multicolumn{2}{c}{Yes} & \multicolumn{2}{c}{Yes} & \multicolumn{2}{c}{No} \\ 

\\[-1.8ex] 
Observations & \multicolumn{2}{c}{3,637} & \multicolumn{2}{c}{3,637} & \multicolumn{2}{c}{5,088} & \multicolumn{2}{c}{3637} \\ 

R$^{2}$ & \multicolumn{2}{c}{0.130} & \multicolumn{2}{c}{0.128} & \multicolumn{2}{c}{0.099} & \multicolumn{2}{c}{0.054} \\ 

Adjusted R$^{2}$ & \multicolumn{2}{c}{0.116} & \multicolumn{2}{c}{0.115} & \multicolumn{2}{c}{0.089} & \multicolumn{2}{c}{0.052} \\ 

\hline 
\hline \\[-1.8ex] 
\multicolumn{9}{l}{\text{$^{***}$p $< 0.01$; $^{**}$p $< 0.05$; $^{*}$p $< 0.1$; standard errors are in parentheses.}} \\  
\end{tabular} 
\end{table*} 

\subsection*{Explaining course influence with VET specificity}

\begin{figure*}[htbp]
  \centering
  \includegraphics[width=1\textwidth]{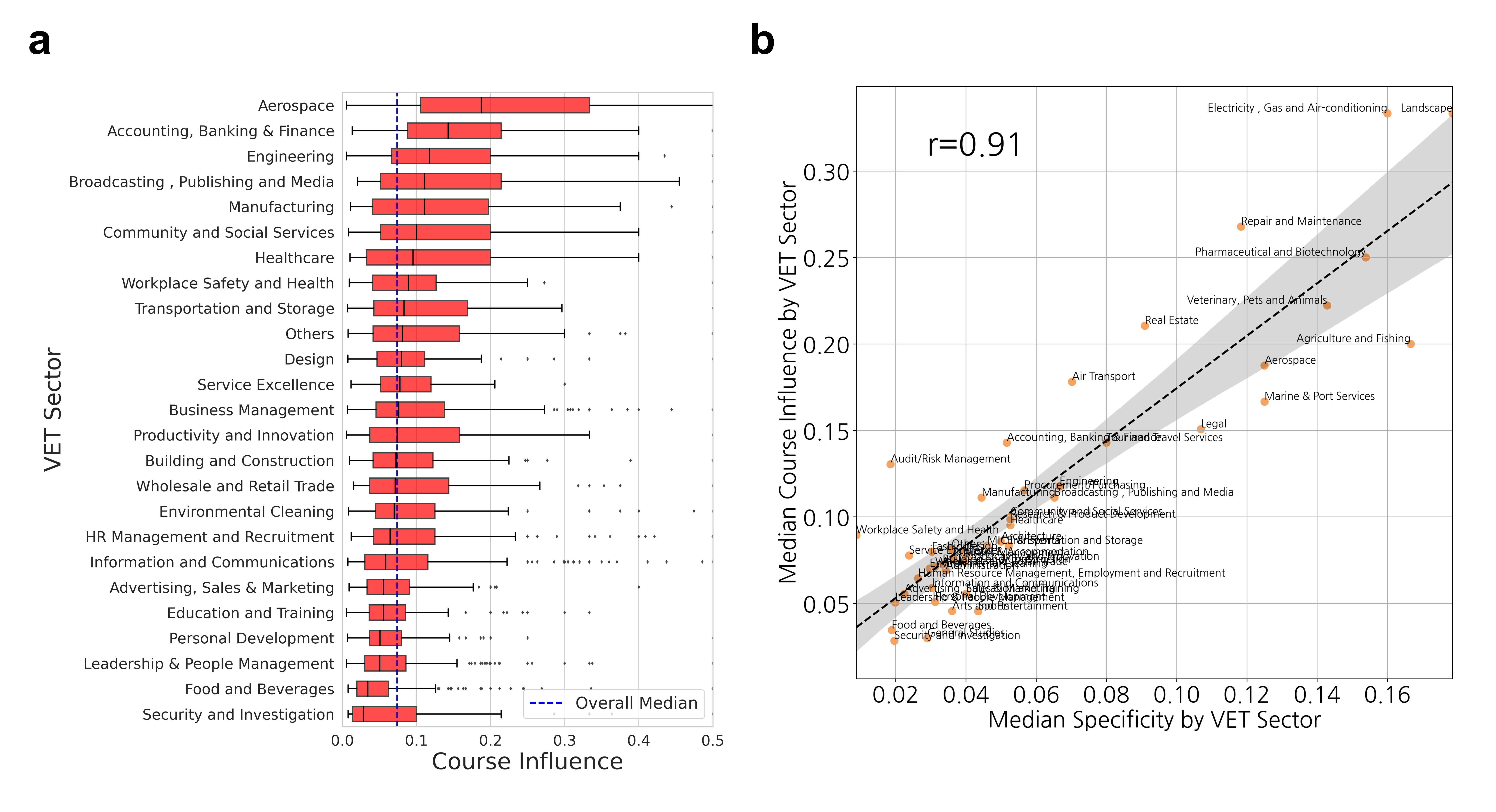}
  \caption{{Course influence and VET specificity.}
  \textbf{a}. Comparison of course influence across different VET sectors. The box plot includes the median, quartiles, and 1.5 interquartile range (IQR) of course influence values for each VET sector. The dashed line indicates the overall median value of course influence over all sectors. \textbf{b}. The relationship between the specificity and course influence at the VET sector level. The horizontal axis represents the median specificity for each VET sector, while the vertical axis represents its median course influence. The text labels and the shade denote the name of VET sectors and the confidence interval of the simple linear regression, respectively.}
  \label{fig3}
\end{figure*}

So far, we focused on the relationship between economic variables and the influence of individual courses. Analyzing the course influence at the level of VET sectors can provide insights into the supply and demand of courses for each sector. Fig.~\ref{fig3} reveals significant differences in the median values of the course influence across various VET sectors. The sectors with relatively high course influence include `Aerospace', `Accounting, Banking and Finance', and `Engineering', while `Food and Beverages', `Education and Training', and `People Management' exhibit low course influence.

This variation in course influence across VET sectors is seemingly associated with each sector’s specificity, i.e., the extent to which a VET course specializes in specific working skill groups, as we observed high influence in sectors such as `Aerospace' and low influence in `Food and Beverages'. To examine this, we define VET specificity by adopting the ubiquity of related skills (see Materials and Methods for details), and compare the median specificity with the median course influence across different VET sectors. Fig.~\ref{fig3}b shows a strong correlation between the specificity and course influence (r=0.91). This finding aligns with the result from the regression analysis, where lower diversity is associated with higher course influence. Both findings imply that when developing new VET courses, specializing in particular target skills as learning objectives can be an effective strategy to maximize the influence.

\subsection*{Course supply by occupation in the Singapore labor market}

The observed imbalance in course influence across VET sectors implies potential oversupply and undersupply of educational opportunities for different occupations in Singapore. To examine the distribution of VET course supply across occupation sectors, we employ an `occupational course supply' (namely, OCS) index which captures the number of courses offered for a particular occupation relative to the average over the entire occupations as:

\renewcommand{\arraystretch}{1.2}  

\begin{equation}
OCS_o = \frac{C_o}{\overline{C}}
\end{equation}
where:
\[
\begin{array}{ll}
    C_o & = \text{courses for occupation } o, \\
    \overline{C} & = \text{average courses per occupation}.
\end{array}
\]

\begin{figure*}[htbp]
  \centering
  \includegraphics[width=1\textwidth]{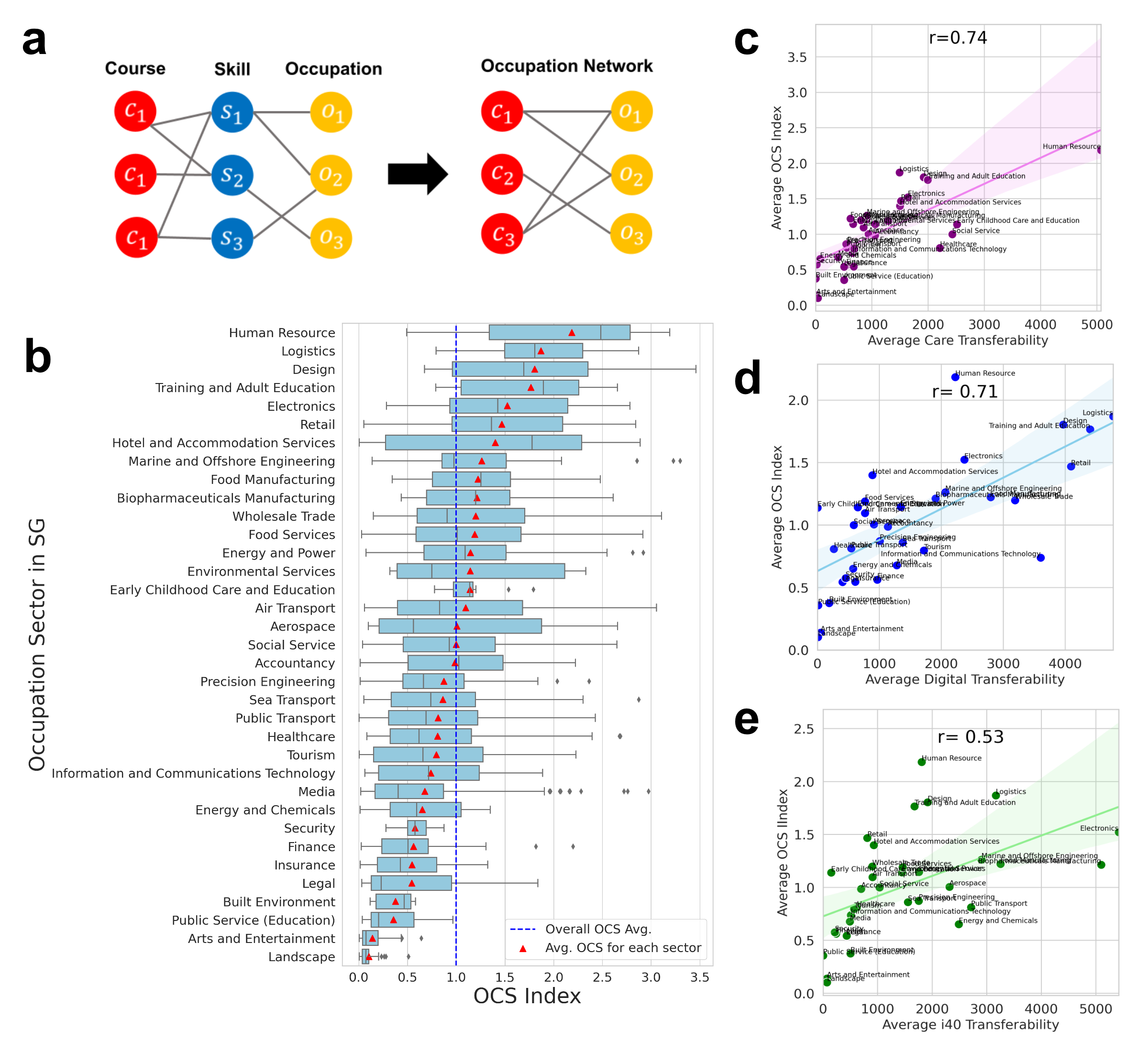}
  \caption{{Occupational course supply and comparison with transferability.} \textbf{a}. Construction of the course-occupation network from the tripartite network of courses, skills, and occupations. Links are connected for entities with embedding similarity higher than 0.6.
  \textbf{b}. OCS index for each occupation sector in Singapore. The average for each sector is denoted by red triangle, and the blue dashed line indicates the average of the OCS index over all occupations. The box plot denotes the median, quartiles, and 1.5 IQR. \textbf{c-e}. Comparison of the average OCS index and average transferability for care economies (c), digital economies (d), and industrial 4.0 economies (e) across occupation sectors. The colored line and shade represent the simple regression plot and its confidence interval, respectively.}
  \label{fig4}
\end{figure*}

We obtained the number of courses offered for each occupation, $C_o$, from the bipartite network of VET courses and occupations as we computed the number of skills covered by a course (i.e., $C_c$) from the network of courses and skills. To get this bipartite network of courses and occupations, we constructed a tripartite network of courses, skills, and occupations by adding the skill-course bipartite network to the course-skill network (see Fig.~\ref{fig4}a for construction of the course-occupation network). We linked skills and occupations in the same way as the course-skill network, utilizing their embedding vectors' textual similarity over 0.6. The course-occupation network was built from the tripartite network by connecting a link between a course and an occupation that can be reached through a skill. Then, we get $C_o$ by counting each occupation's degree, i.e., the number of courses linked to the occupation, on the bipartite network.

The observed OCS index shows that occupational sectors such as HR, logistics, and design tend to have more courses, while other sectors like Built Environment, Public Service, Arts and Entertainment, and Landscape have relatively fewer courses. This suggests that occupation sectors with lower OCS indices less benefit from the SkillsFuture program with fewer courses and may require more VET course supply.

Furthermore, we elucidated the relationship between course supply and future economies by taking the transferability values for each occupation sector. The occupational transferability $T_{o,e}$ for occupation $o$ and future economy $e$ was obtained by taking the average of the skill transferability $T_{s,e}$ over the skills $s$ linked to occupation $o$ as $T_{o,e}=\sum_{s}A_{os}T_{s,e}$, where $A_{os}$ is the adjacency matrix of the occupation-skill bipartite network. By doing so, we can assess how skills obtained from various courses are related to occupational transferability in the context of the future economy. 

As a result, the OCS index and occupational transferability at the sector level show the highest correlation for care economies ($r=0.74$), followed by the digital ($r=0.71$) and the fourth industrial revolution ($r=0.53$) economies (see Fig.~\ref{fig4}c, ~\ref{fig4}d, and ~\ref{fig4}e). The green transferability is not significantly correlated with the OCS index, despite the growing importance of sustainable development, demonstrating that the VET market has not overly saturated this sector with training courses. As such, the SkillsFuture VET market exhibits a disproportionate concentration in particular jobs.

\section*{Discussion}

To assess the influence of Vocational Education and Training (VET) in Singapore’s labor market, this study utilized a course/skill/occupation network based on their textual similarity. Leveraging the constructed network, we quantified the course influence of each VET course and found that course sectors covering specific skills tended to have high course influence, with the suggested specificity index. Lastly, we demonstrated the imbalance of educational opportunities for different occupations, denoted by a high course supply for occupations with general skills and in care economies.

The use of text data in this study is particularly noteworthy due to its high accessibility as a method for quantification. This approach makes it possible to quantify some features that had not been previously quantified, despite the availability of relevant text information. We expect this approach to help address data challenges in evaluating the effectiveness of VET programs. Analyzing previously untapped text data could open new avenues for education research, providing a deeper understanding of VET's impact on skills development and labor market outcomes. Also, this method is not limited to VET; it can be applied to a wide range of education studies, offering useful insights into how different forms of education can influence workforce readiness and skill acquisition.

The indices proposed in this study could offer valuable contributions to existing research in several aspects. Firstly, to the authors' best knowledge, this study appears to be among the first to holistically and quantitatively explore the complex relationships between VET programs, skills, and occupations. By doing so, it seeks to extend the scope of labor market research beyond the traditional focus on the relationship between specific skills and employment. Moreover, the study introduces a new way of measuring the effectiveness of vocational education, which has predominantly relied on qualitative research based on surveys and course reviews. Lastly, the study contributes to the context of VET evaluation by proposing the VET specificity index adopted from the ubiquity index which has been originally used in the field of economic complexity. 

Given the evolving landscape of future regional economies and work skills influenced by climate change, digital transformation, and pandemics, there is a growing demand for vocational education to adapt to such changes~\citep{mcgrath2016skills, ostmeier2022building}. For example, during the COVID-19 pandemic, a significant increase in remote or hybrid work arrangements led to a fundamental shift in working styles, necessitating new skills and tools for existing workers~\citep{yang2022effects}. As a result, new vocational education and training (VET) courses have emerged to address these changing demands. In this light, the methods and indices proposed by this study facilitate assessing the macroeconomic and social influences of VET, aiding policymakers in making informed decisions in different scenarios.

Lastly, this study is not without limitations. One limitation is its sole focus on the four future economies, without considering the entirety of occupations or industries due to the lack of data. Incorporating a more comprehensive dataset of 
employment would allow a more refined measurement of the VET influence in Singapore and provide a more holistic understanding of its effectiveness and contribution to Singapore's workforce and economy. Another constraint would be the bias in the data stemming from the locality of Singapore being a city-state that possesses unique characteristics in terms of the industrial landscape and employment-population structure. This uniqueness makes it challenging to generalize the analysis to other countries or regions. One potential direction for future research for mitigating this locality is to compare the inequality of VET at different regional levels (i.e., urban vs. rural) or the country level (i.e., developed countries vs. developing countries). We expect this comparative analysis could provide valuable insights into the effectiveness and accessibility of VET across different socio-economic contexts~\citep{feng2023dynamic}.



\section*{Materials and Methods}

\subsection*{SkillsFuture dataset}
The study uses textual data on the learning objectives and course descriptions of Vocational courses (n=13,695), along with the information on skills (n=2,580) and occupations (n=1,456), sourced from SkillsFuture Singapore~\cite{Skillsfuture}. SkillsFuture is a national initiative designed to empower Singaporeans with opportunities for skill development and lifelong learning~\citep{fung2020developing}. As part of the investment in human capital, all Singapore citizens aged 25 and above are entitled to receive SkillsFuture credits, which can be used to cover diverse skills-related courses available at~\url{https://www.myskillsfuture.gov.sg}. The portal offers courses across various training areas associated with major national industries in Singapore. Comprehensive information about their curricula currently in service, including details about training agencies and course objectives, is accessible through the SkillsFuture API. While the portal provides detailed information about individual SkillsFuture programs, the study utilized course categories, skill categories, skill transferability for 4 distinct future economy sectors, and occupation categories.

\subsection*{\texorpdfstring{Constructing the course network \\ based on text similarity}{Constructing the course network based on text similarity}}

We first built a bipartite network of VET courses and skills by connecting VET courses with skills based on their calculated textual similarities to create a monopartite VET course network ultimately. This involved embedding text descriptions of each entity into 384-dimensional vectors using the pre-trained Sentence Transformer model without fine-tuning~\citep{reimers2019sentence} as in Fig.~\ref{fig1}a. Specifically, we utilized the ``all-MiniLM-L6-v2'' model, which is optimized for downstream tasks such as information retrieval, clustering, and sentence similarity measurement. 

 We encoded inputs with a maximum of 200-word tokens, padding shorter ones and truncating longer ones. The linkage between a VET course and a work skill was established by computing their cosine similarity with a threshold of 0.6. For the bipartite network of courses and skills, the adjacency matrix \( A_{cs} \) takes a value of 1 when a course $c$ effectively covers skill $s$, and 0 otherwise as
\begin{equation}
\label{eq:formula1}
A_{cs} = \begin{cases}
1 & \text{if } \frac{\bm{V}_c \cdot \bm{V}_s}{\Vert\bm{V}_c\Vert\Vert\bm{V}_s\Vert} \geq 0.6,\\
0 & \text{otherwise},
\end{cases}
\end{equation}
where $\bm{V}_c$ and $\bm{V}_s$ are the embedding vectors of course $c$ and skill $s$, respectively. We set this threshold of 0.6 by taking the inflection point in the function of the giant component fraction and threshold (see Fig~\ref{S3_Fig} for the function of the giant component fraction and threshold). Subsequently, the resulting bipartite network was projected to generate a VET course network, where nodes are VET courses and links are connected between two VET courses that are connected to the same skills in the bipartite network (see Fig.~\ref{fig1}b).

\subsection*{VET specificity}
The proposed index, VET specificity, serves as a proxy to evaluate the extent to which VET courses within the network specialize in specific skills, as adopted from the studies on economic complexity~\citep{hidalgo2009building, hidalgo2007product, hausmann2011network}. 
Leveraging \( A_{cs} \), the number of occupations and skills covered by each course (i.e. diversification), \( k_{c,0} \), and the number of different courses that cover a particular skill (i.e. ubiquity), \( k_{s,0} \), can be computed as $k_{c,0} = \sum_{s}{A_{cs}}$ and $k_{s,0} = \sum_{c}{A_{cs}}$, respectively.
Then, \( k_{c,1} \) represents course $c$'s average ubiquity of the connected skills as
\begin{equation}
\label{eq:formula3}
k_{c,1} = \frac{1}{N_c} \sum_{s} A_{cs}k_{s,0},
\end{equation}
where $N_c$ is the number of skills connected to course $s$. Using this, we define VET specificity as the inverse of the average ubiquity as $S_{c} = 1/k_{c,1}$ which measures how much a course is connected to less ubiquitous (i.e., more specific) skills.

\subsection*{VET transferability on future economy}
Transferable skills are abilities that are applicable and useful in various situations and aspects of life~\citep{nagele2017competence}. In the context of Skillsfuture dataset, transferability refers to the number of occupations that demand skills related to four distinct future economy sectors of green economy~\citep{georgeson2017global}, digital economy~\citep{carlsson2004digital}, care economy~\citep{folbre2018developing}, and the fourth Industrial Revolution economy provided by SkillsFuture~\citep{schwab2017fourth}. We computed the transferability $T_{c,e}$ for each VET course node $c$ and each future economy sector $e$ by aggregating the transferability values of every skill node neighbor connected to individual course nodes within the constructed VET network as
\begin{eqnarray}
\label{eq:formula4}
    T_{c,e} &=& \sum_{s}{A_{cs}T_{s,e}},
\end{eqnarray}
where $T_{s,e}$ is the transferability value of skill $s$ for future economy sector $e$, provided by the SkillsFuture dataset. A higher transferability value for a specific course node implies that the course is more likely to contribute to generating more employment related to that particular future economy.

\subsection*{VET skill diversity}
The diversity of each VET course is expressed through the Shannon entropy value for skill categories connected to each course node. The entropy value of a course gets higher when the categories of skills connected to the course are more evenly distributed. Here the skill diversity $H_{c}$ of course $c$ is defined as
\begin{eqnarray}
\label{eq:formula5}
H_{c} &=& -\sum_{x} p_{c}(x) \log p_{c}(x),
\end{eqnarray}
where \( p_{c}(x) \) represents the ratio of the number of skills in skill category $x$ to the total number of skills connected to course $c$.

\begin{acknowledgments}
I.H. was supported by the National Research Foundation of Korea (NRF) grant funded by the Korean government (MSIT) (RS-2023-00278682, RS-2023-00242528).
\end{acknowledgments}

\nocite{*}
\bibliographystyle{apsrev4-2}
\bibliography{references}%

\clearpage 

\section*{Supporting information}

\renewcommand{\thefigure}{S\arabic{figure}}  
\setcounter{figure}{0}  

\begin{figure}[htbp]
    \centering
\includegraphics[width=0.45\textwidth]{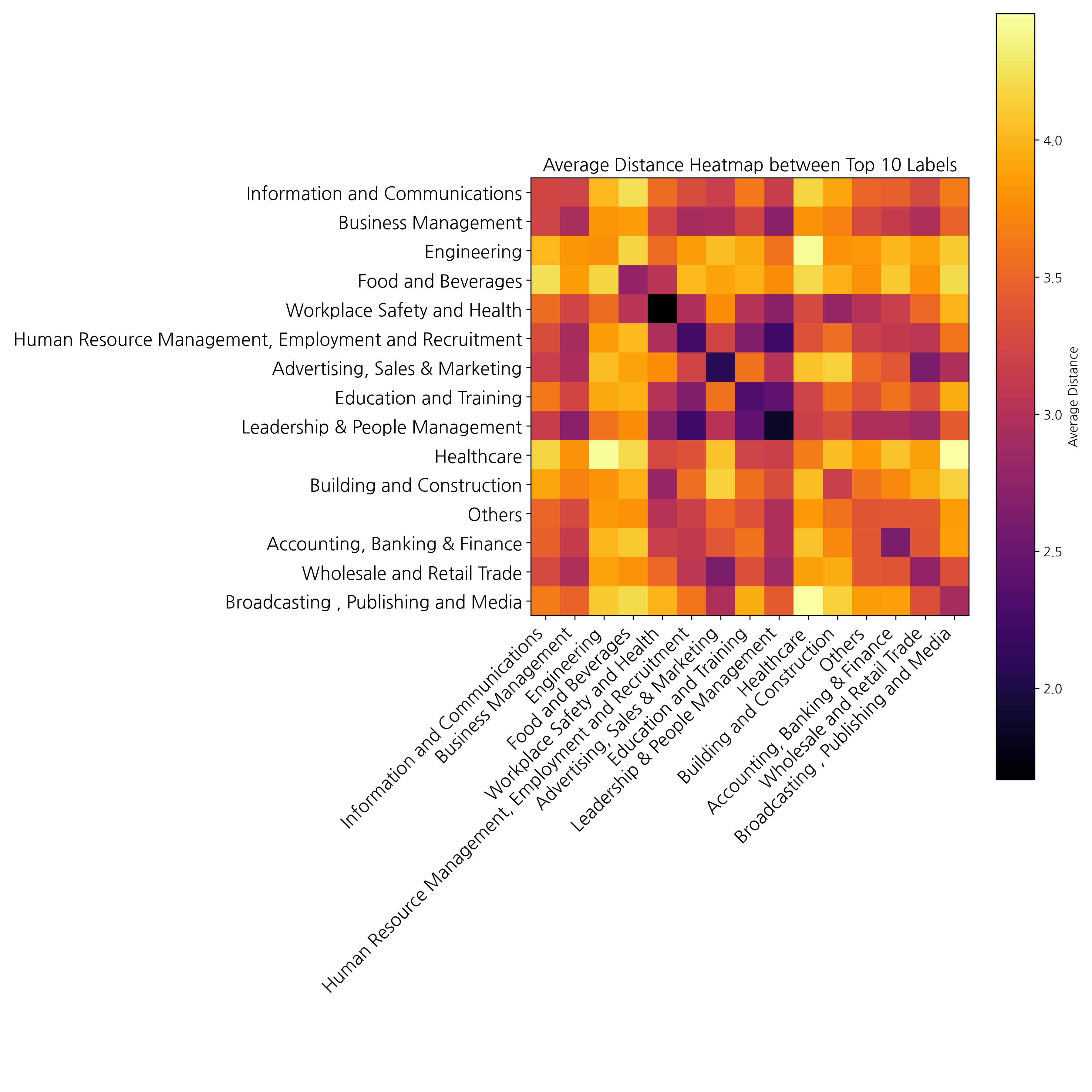} 
    \caption{{Average path length between VET sectors within the course network.} The heatmap demonstrates the average path length between the top 15 VET sectors. The lighter the plot, the farther the distance between the sectors.}
    \label{S1_Fig} 
\end{figure}

\begin{figure}[htbp]
    \centering
\includegraphics[width=0.45\textwidth]{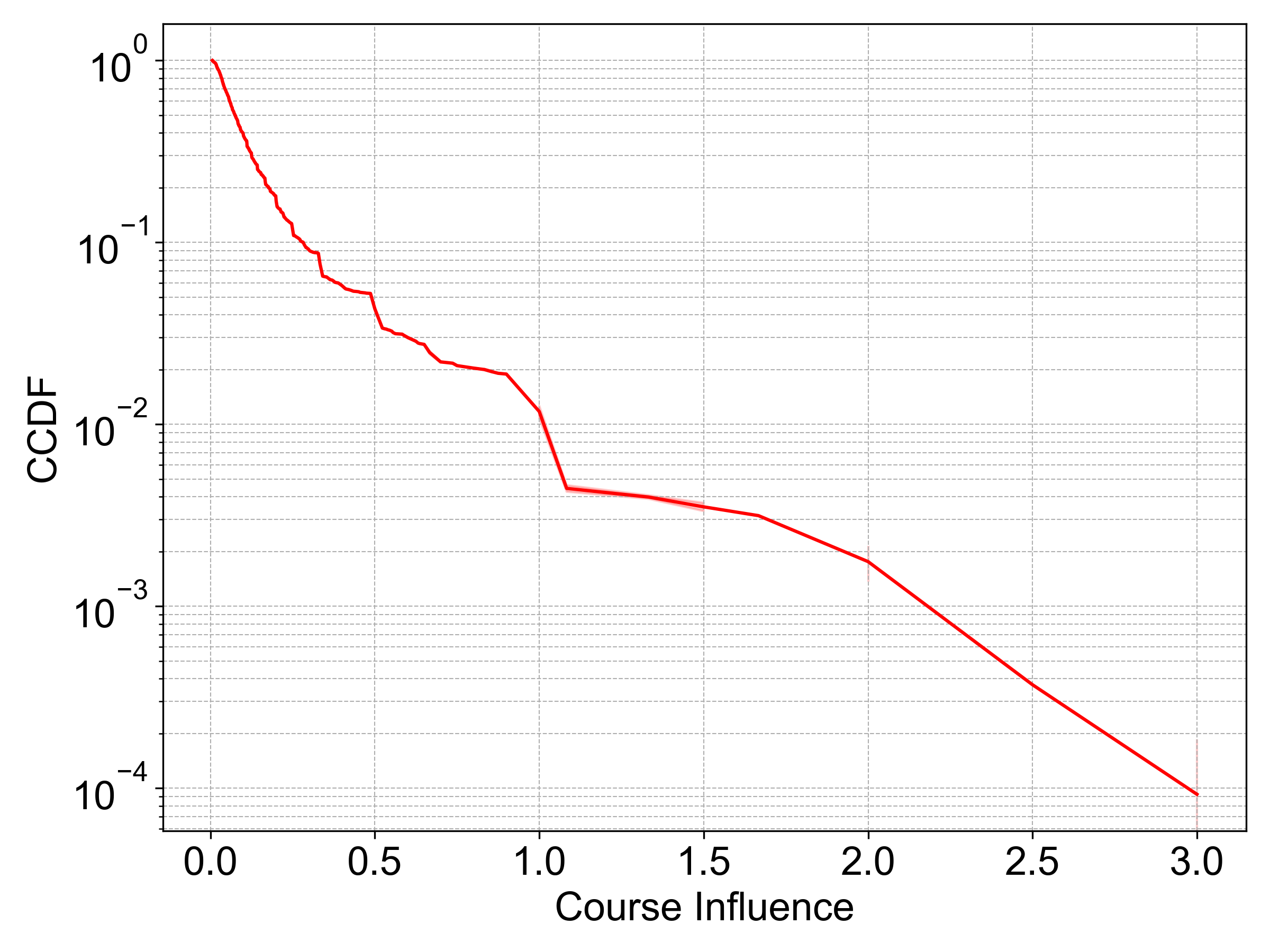}  
    \caption{{Distribution of Course Influence.} The figure displays a complementary cumulative distribution function (CCDF) plot of the course influence values. The plot demonstrates that the original course influence data follows an exponential distribution.} 
    \label{S2_Fig}  
\end{figure}

\begin{figure}[htbp]
    \centering
\includegraphics[width=0.45\textwidth]{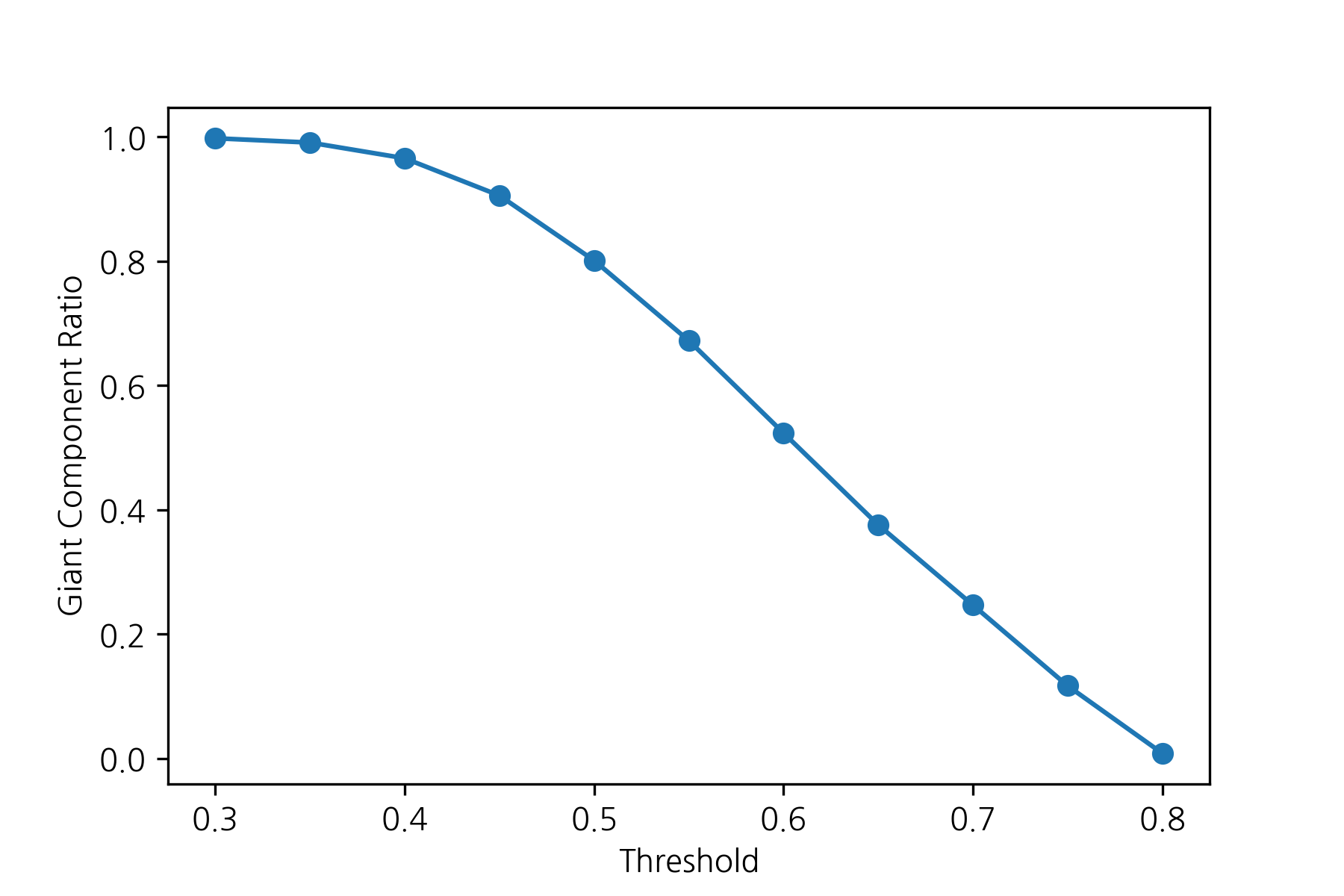}  
    \caption{{Giant component fraction as a function of the similarity threshold to link entities.} The plot demonstrates the ratio of nodes belonging to the giant component of the course network for each threshold. The course network was made by projecting the course-skill bipartite network where the links are connected for the text similarity higher than the given threshold.}
    \label{S3_Fig}  
\end{figure}

\end{document}